\documentclass[runningheads]{llncs}

\usepackage{graphicx}
\usepackage{bm}
\usepackage{caption}

\usepackage{graphicx}
\usepackage{multirow}	
\usepackage{arydshln}   
\usepackage{color}		
\usepackage{marvosym}	
\usepackage{subfigure}  

\usepackage{float}      
\usepackage{amsmath}    
\usepackage{algorithmic}
\usepackage{algorithm}  
\usepackage{indentfirst}

\usepackage{booktabs}
\usepackage{array}
\usepackage{balance} 
\usepackage{multirow}
\usepackage[normalem]{ulem}
\usepackage{color}
\definecolor{lightgray}{RGB}{221,235,247}
\usepackage{colortbl}  
\usepackage{xcolor}
\useunder{\uline}{\ul}{}
\usepackage{subfigure}

\usepackage{cite}

\usepackage[colorlinks,linkcolor=red,anchorcolor=blue,citecolor=black]{hyperref}

\begin{document}

\title{
    CherryRec: Enhancing News Recommendation Quality via LLM-driven Framework
}

\titlerunning{CherryRec}



\author{
	Shaohuang Wang \and
	Lun Wang  \and
        Yunhan Bu \and
        Tianwei Huang
}
\institute{Xinjiang University, Urumqi, China    \\\email{\{107473, 107243, 10745, 10764\}@stu.xju.edu.cn}}
\maketitle              

\begin{abstract}

Large Language Models (LLMs) have achieved remarkable progress in language understanding and generation. Custom LLMs leveraging textual features have been applied to recommendation systems, demonstrating improvements across various recommendation scenarios. However, most existing methods perform untrained recommendation based on pre-trained knowledge (e.g., movie recommendation), and the auto-regressive generation of LLMs leads to slow inference speeds, making them less effective in real-time recommendations.To address this, we propose a framework for news recommendation using LLMs, named \textit{CherryRec}, which ensures the quality of recommendations while accelerating the recommendation process. Specifically, we employ a Knowledge-aware News Rapid Selector to retrieve candidate options based on the user's interaction history. The history and retrieved items are then input as text into a fine-tuned LLM, the Content-aware News Llm Evaluator, designed to enhance news recommendation capabilities. Finally, the Value-aware News Scorer integrates the scores to compute the CherryRec Score, which serves as the basis for the final recommendation.We validate the effectiveness of the proposed framework by comparing it with state-of-the-art baseline methods on benchmark datasets. Our experimental results consistently show that CherryRec outperforms the baselines in both recommendation performance and efficiency.The project resource can be accessed at: \url{https://github.com/xxxxxx}
\end{abstract}

\section{Introduction}
\noindent The advent of Large Language Models (LLMs) \cite{zhao2024llm, lin2024data, peng2023rwkv, touvron2023llama} has heralded a new era in the field of artificial intelligence, particularly in tasks involving language understanding and generation. With the ability to process vast amounts of textual data \cite{orvieto2023resurrecting, sun2023retentive}, LLMs have become pivotal in enhancing the capabilities of recommendation systems, which are the cornerstone of personalized content delivery in various domains such as e-commerce, entertainment, and news aggregation.Despite the significant strides made by LLMs, their application in real-time recommendation systems has been hindered by several challenges \cite{chen2023palr,  wang2023zero, zhang2023recommendation}. The auto-regressive nature of LLMs results in slow inference times, which is a critical drawback for systems requiring rapid and responsive recommendations. Furthermore, the majority of existing methods rely on pre-trained knowledge for untrained recommendations, such as movie suggestions, which may not be optimal for dynamic and user-specific content like news articles.

To bridge this gap, we introduce CherryRec, a novel framework for news recommendation that leverages the power of LLMs while addressing the limitations of current approaches. CherryRec is designed with a dual focus on the quality and speed of recommendations. It streamlines the recommendation process with a Knowledge-aware News Rapid Selector, pinpointing relevant news candidates from extensive datasets by analyzing user interactions and content attributes. These candidates are then subjected to the scrutiny of the Content-aware News Llm Evaluator, a specialized LLM finely tuned to discern user preferences and contextual cues, thereby enriching the personalization of recommendations.The culmination of this process is the Value-aware News Scorer, which amalgamates insights to formulate the CherryRec Score. This metric encapsulates the personalized value of news items, ensuring that recommendations are timely, pertinent, and tailored to user interests.

In this paper, we delve into the methodology behind each component of CherryRec and demonstrate its efficacy through rigorous experimentation on benchmark datasets. We compare CherryRec's performance with state-of-the-art baseline methods and present empirical evidence that underscores the superiority of our approach in terms of recommendation accuracy and efficiency.

\par The contributions of our work can be summarized as follows:
\begin{enumerate}
    \item We proposed CherryRec, a novel framework for news recommendation that leverages the power of LLMs while addressing the limitations of current approaches.This framework not only quickly filters out low-value news but also recommends high-quality news by deeply understanding user preferences and integrates multi-dimensional scores to generate the final recommendation sequence.
    \item We designed a new metric, CherryRec Score for Comprehensive News Evaluation. It enables a more accurate and personalized news recommendation by combining insights from multiple modules, ensuring that the recommended news items are both timely and relevant to the user’s interests.
    \item We validated the Effectiveness of CherryRec on Different Datasets. Experimental results on benchmark datasets such as MIND, Yahoo R6B, and Adressa demonstrate that CherryRec outperforms existing state-of-the-art methods in both recommendation performance and efficiency.
\end{enumerate}

\section{Related Work}
\noindent The development of LLM  has made significant strides, with the introduction of ChatGPT being a notable example\cite{liu2023chatgpt,gao2023chat,sun2023chatgpt,petrov2023generative}. This powerful chatbot based on LLMs is a testament to the advancements in the field. The progress of these LLMs can be largely attributed to two main factors: (1) the expansion of the scale of language models, and (2) the extension of the text corpus during the pre-training phase. LLMs have been applied to recommendation systems to understand the textual characteristics of items and enhance recommendation performance\cite{Cui2022M6RecGP,kang2023llms,Li2022PersonalizedPL,cui2021disentangled}. Most existing LLM-based recommendation systems are unoptimized, using pre-trained knowledge to generate recommendations for the next item. For example, Chat-REC utilizes ChatGPT to understand user preferences and improve interactive and interpretable recommendations.

\section{Problem Formulation}

\noindent For the task in this paper, we can proceed with mathematical modeling and symbolic representation in the following manner:

Firstly, let's define some basic symbols and sets: \( U \): The set of users, \( U = \{u_1, u_2, ..., u_M\} \). \( V \): The set of news items, \( V = \{v_1, v_2, ..., v_N\} \). \( S_u \): The historical interaction sequence of user \( u \), \( S_u = \langle v_{u1}, v_{u2}, ..., v_{ut}, ..., v_{u|S_u|} \rangle \), where \( |S_u| \) denotes the length of the sequence.

Next, we define the recommendation task: Input: The collection of historical interaction sequences for all users \( S = \{S_{u_1}, S_{u_2}, ..., S_{u_M}\} \). Output: For any user \( u \), at time \( t \), the next likely interactive news item \( \hat{v}_{u,t+1} \).

We can model the news recommendation problem as a prediction problem, that is, predicting which news item \( v \) user \( u \) will interact with at the next time point \( t+1 \) given the historical interaction sequence \( S_u \). This can be represented by conditional probability as follows:

\[ \hat{v}*{u,t+1} = \arg\max*{v \in V} P(v | S_u, \theta) \]

where \( \theta \) represents the model parameters, and \( P(v | S_u, \theta) \) represents the probability of news item \( v \) being interacted with by user \( u \) given the user's history \( S_u \) and model parameters \( \theta \).

\section{Methodology}
\begin{figure}[H]
	\begin{minipage}[t]{1\textwidth}	
	\centering
		\includegraphics[width=0.96\textwidth]{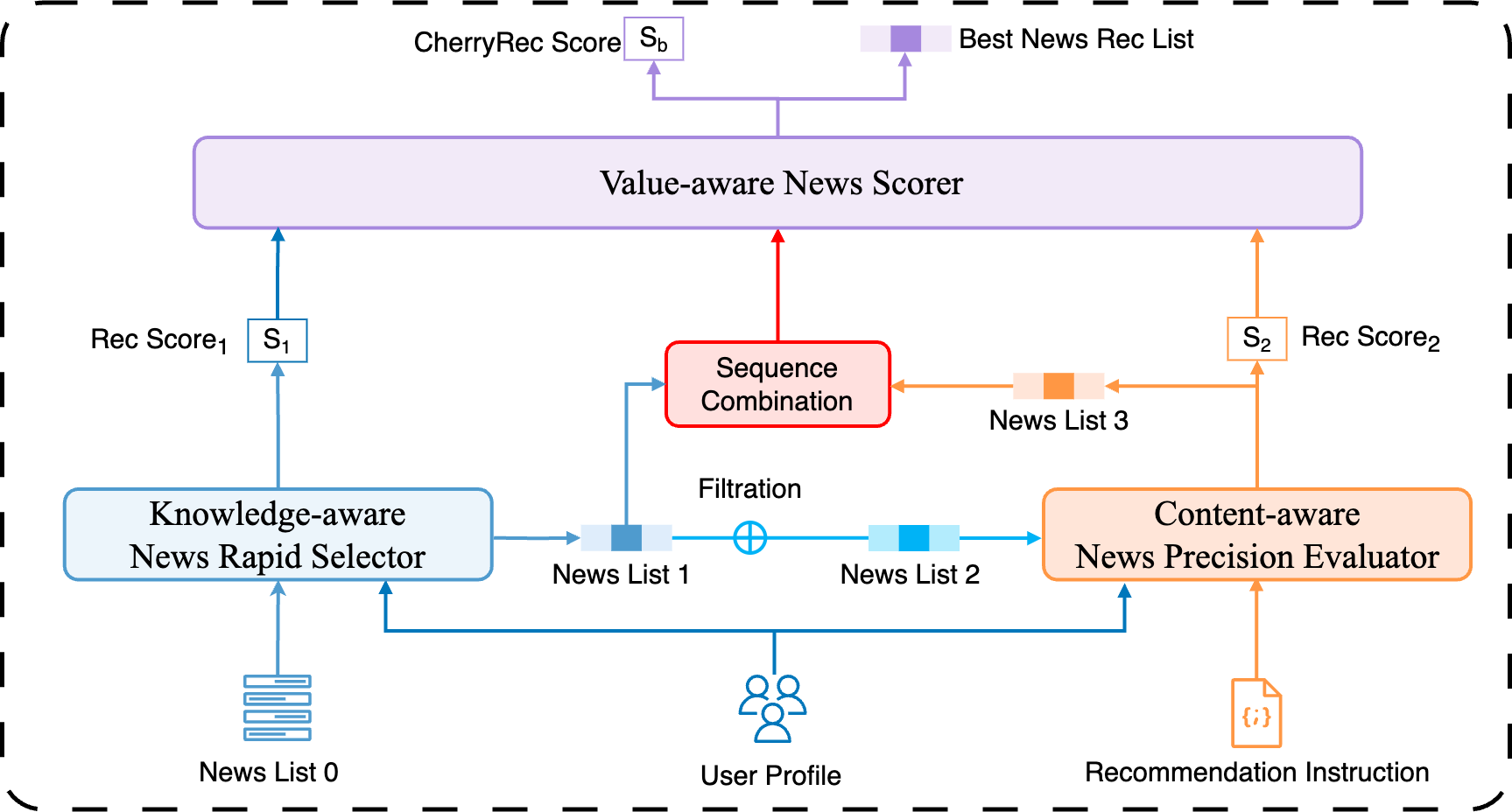}
		\caption{The overall framework of CherryRec. Here has three key components: the Knowledge-aware News Rapid Selector (KnRS) for filtering news, the Content-aware News LLM Evaluator (CnLE) for understanding user preferences, and the Value-aware News Scorer (VaNS) for synthesizing a personalized recommendation score.}	
		\label{fig-prompt-prob}
	\end{minipage}	
\end{figure}

\noindent In this section, we introduce \textit{CherryRec} for news recommendation which can consider both the personal interest of users and the popularity of candidate news.
First, we introduce the overall framework of \textit{CherryRec}, as shown in Fig.~\ref{fig-prompt-prob}.

\subsection{Knowledge-aware News Rapid Selector (KnRS)}

\noindent In this section, we designed the Knowledge-aware News Rapid Selector with the purpose of quickly filtering out low-quality news through a news value assessment model. This module takes user profiles and the news sequence to be recommended as input, and quickly filters out 95\% of low-value news by integrating a news value assessment model that combines five news value metrics: user relevance, source credibility, news timeliness, online attention, and content novelty, with detailed in Fig.~\ref{4.2}.

The KnRS \cite{hou2023large, liu2023chatgpt,liu2023llmrec,li2023prompt,peng2023rwkv} constructs a news value assessment model through multidimensional feature fusion technology. Its neural network architecture can integrate these data and understand them, ultimately generating a comprehensive score to measure the five news value characteristics and filter out low-value news. It combining the User Relevance(\( f_1 \)), Source Credibility (\( f_2 \)), News Timeliness (\( f_3 \)), Online Attention (\( f_4 \)) and Content Novelty (\( f_5 \)) indicators, the news value assessment model \( V \) of KnRS can be expressed as:
\[ V(u, N) = w_1 \cdot f_1(u, N) + w_2 \cdot f_2(N) + w_3 \cdot f_3(N) + w_4 \cdot f_4(H) + w_5 \cdot f_5(N) \]
where \( w_1, w_2, w_3, w_4, w_5 \) are weight coefficients used to adjust the importance of different indicators in the news value assessment.

KnRS also builds a user profile library for each user, which is necessary for high-quality recommendations, including role positioning, field focus, collection list, and behavioral information.

\subsection{Content-aware News LLM Evaluator (CnLE)}

\begin{figure}[H]
  \centering
  \subfigure[Content-aware News Llm Evaluator]{\includegraphics[width=0.49\linewidth]{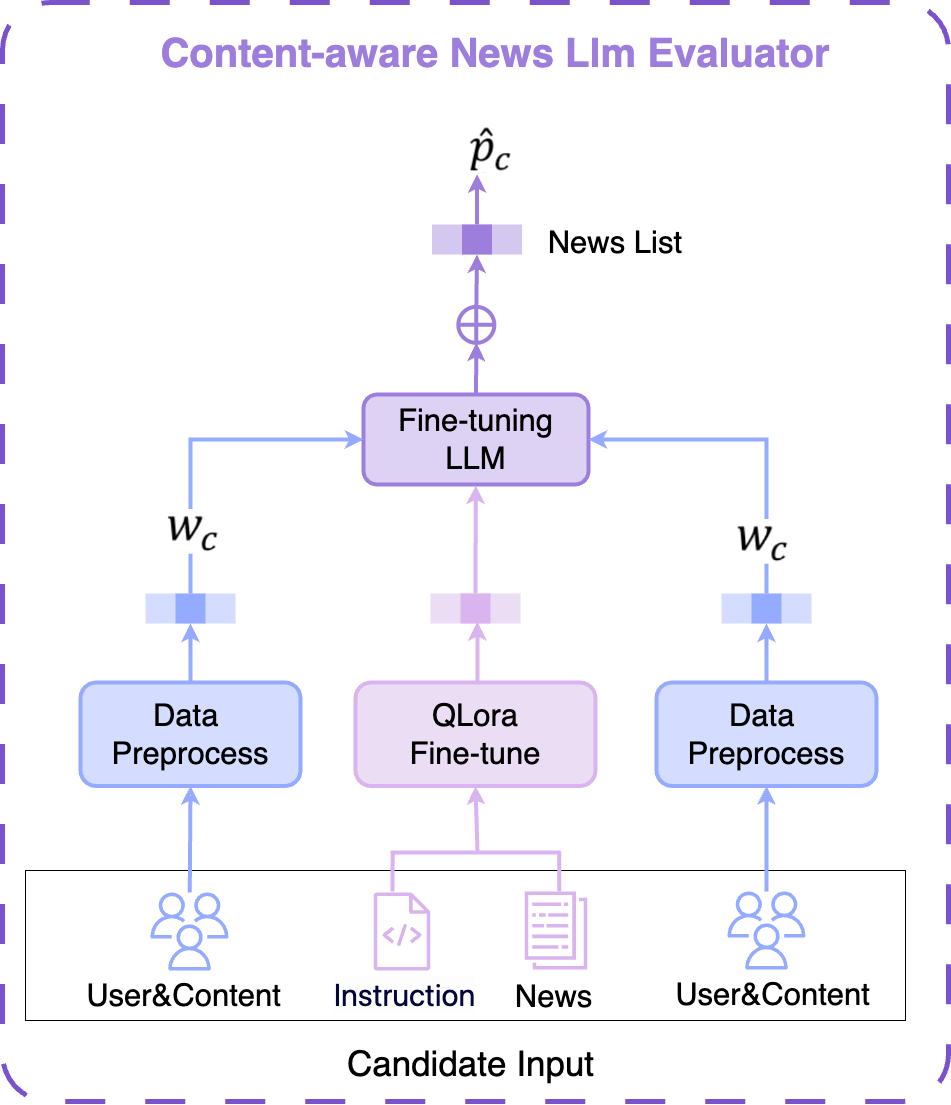}}\label{4.2(a)}
  \subfigure[Knowledge-aware News Selector]{\includegraphics[width=0.49\linewidth]{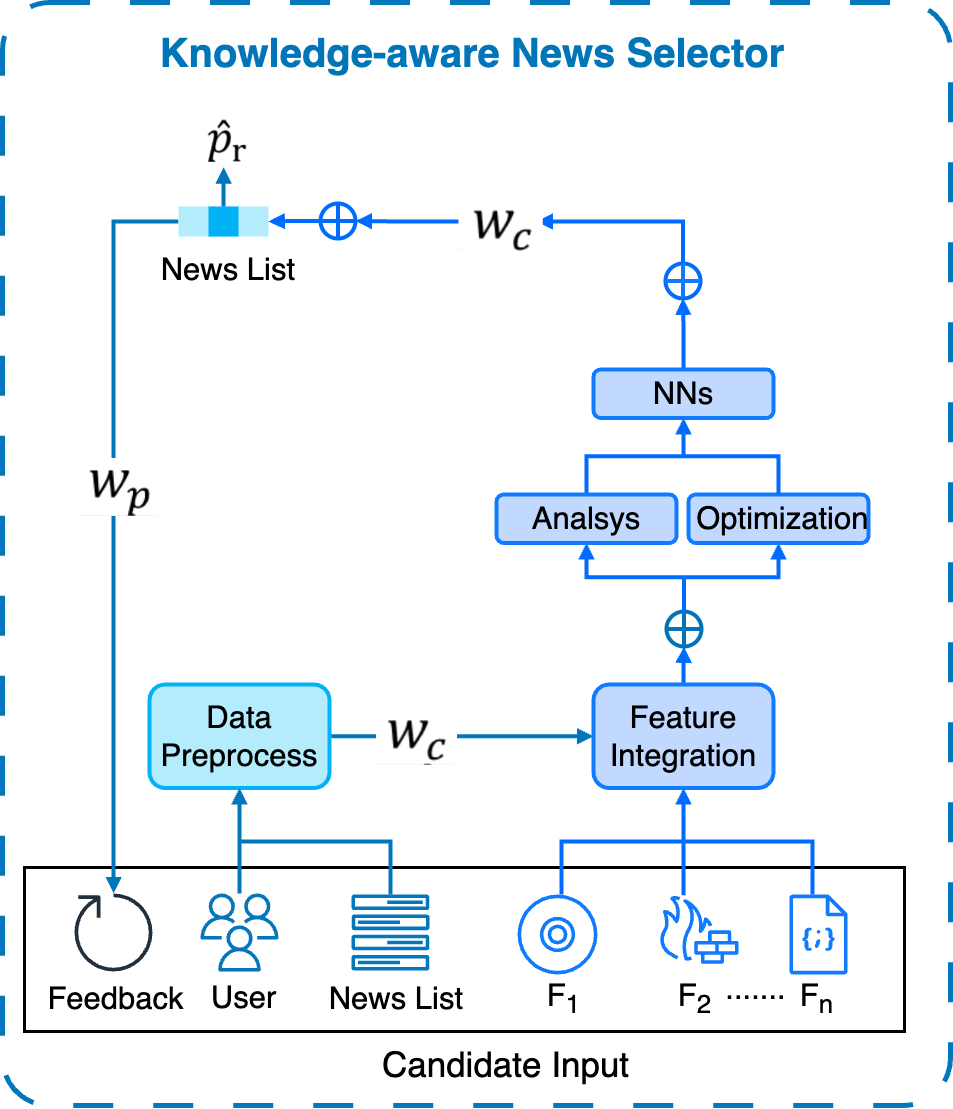}}\label{4.2(b)}
  
  \caption{Knowledge-aware News Rapid Selector (KnRS) quickly identifies relevant news candidates by assessing user interaction history and content attributes. Content-aware News LLM Evaluator (CnLE) refines selections using a fine-tuned LLM, deeply understanding user preferences to enhance personalized news recommendations.}
  \label{4.2}
\end{figure}

\noindent In this section, we designed the Content-aware News LLM Evaluator with the purpose of recommending high-preference news by deeply understanding user preferences through the LLM. This module takes the Prompt, the news sequence to be recommended, and the user profile as input, and uses an LLM fine-tuned for various recommendation task types to filter out the high-quality news that the user is most interested in, with detailed in Fig.~\ref{4.2}.

The specific design of the model is as follows: We selected Qwen2-7B \cite{touvron2023llama} as the base LLM. To meet the different recommendation task requirements reflected by different user profiles, we constructed a dataset covering various recommendation task types. In three different datasets, we annotated four different types of recommendation tasks, totaling 323 prompt templates. Each prompt template includes three fields: prompt, news sequence to be recommended, and user profile. Subsequently, we fine-tuned the base LLM, optimizing the cross-entropy loss on the model's output tokens to ensure that the generated text is related to the recommendation task. Through the above steps, we achieved a high-quality news recommender that can accurately recommend the most relevant news from the candidate pool.

\subsection{Value-aware News Scorer (VaNS)}

\noindent In this section, we introduce the Value-aware News Scorer (VaNS), designed to synthesize the CherryRec score for news recommendation. VaNS inputs the sequences from the Knowledge-aware News Selector (KnRS) and Content-aware News LLM Evaluator (CnLE), calculating the CherryRec score to produce the final news sequence. The metric's strength is its integration of assessments from multiple modules, considering dimensions like relevance, novelty, and user preferences for precise recommendations. It also balances module strengths and limitations for swift and accurate recommendations, enhancing model robustness and reducing reliance on single data sources or biases. It allows for fine-tuning of the recommendation system for improved performance in specific domains.

\begin{algorithm}[H]
    \caption{Procedure of Value-aware News Scorer (VaNS)}
    \label{alg:VaNS}
    \renewcommand{\algorithmicrequire}{\textbf{Input:}}
    \renewcommand{\algorithmicensure}{\textbf{Output:}}
    \begin{algorithmic}[1]
        \REQUIRE Sequences from KnRS and CnLE, parameters $\lambda, \alpha, \beta, \gamma$.
        \ENSURE Final recommended news sequence $C$.
        
        \STATE Standardize input sequences from KnRS and CnLE.
        
        \STATE Initialize model parameters $\beta = \{\beta_0, \beta_i, \beta_{ij}\}$.
        
        \FORALL{$t \in \{1, \ldots, T\}$} 
            \STATE Extract multi-dimensional features for the $t$-th news item. Then compute the polynomial regression prediction for the $t$-th item using the current model parameters:
            \[
            \hat{y}_t = \beta_0 + \sum_{i=1}^{n} \beta_i A_{it} + \sum_{i<j} \beta_{ij} A_{it} A_{jt} + \epsilon_t
            \]
            
            \STATE Update model parameters $\beta$ using gradient descent with L2 regularization:
            \[
            \beta \leftarrow \beta - \eta \nabla_{\beta} \left\{ \sum_{i=1}^{m} (\hat{y}_i - y_i)^2 + \lambda \sum_{j=1}^{p} \beta_j^2 \right\}
            \]
            \COMMENT{$\eta$ is the learning rate.}
        \ENDFOR
        
        \STATE Employ an ensemble of models with weighted fusion and rank news items based on the ensemble prediction:
\[
\hat{y}_{\text{final}} = \alpha \hat{y}_{\text{lr}} + \beta \hat{y}_{\text{rf}} + \gamma \hat{y}_{\text{svm}}, \quad \text{Rank}(i) = \text{argsort}(\hat{y}_{\text{final}})
\]
        
        \STATE Assemble the recommended news sequence $C$ from the ranked items:
        \[
        C = \text{News\_Items}[\text{Rank}]
        \]
        
        \RETURN $C$
    \end{algorithmic}
\end{algorithm}

\section{Experiment}
\noindent In this section,  we will present the detailed experiment settings and the experimental results to demonstrate the effectiveness of CherryRec. We conduct experiments to answer the following three research questions (RQ) about CherryRec:

\begin{itemize}
\item \textbf{RQ1}: Could CherryRec outperform the baseline methods and achieve state-of-the-art results in News recommendation task?
\item \textbf{RQ2}: Is the design of each module in CherryRec reasonable, and is there a pattern to the selection of hyperparameters?
\item \textbf{RQ3}: Is there a more intuitive explanation to illustrate the performance of CherryRec?
\end{itemize}

\begin{table*}[t]
    \centering
    \caption{Instruction tuning data with different rec tasks fine-tuned by  Qwen2-7B.}
    \scalebox{1}{
    \begin{tabular}{c}\toprule
    \begin{minipage}{\linewidth}
    \begin{small}
    \underline{Task:} Recommendation based on User Interest and Role Positioning\\
    \underline{Instruction:}  Recommend a news item and its ID from the candidate news list according to the user's interests and responsibilities.
    \\ \underline{Input:} User's interest and work tasks; The news waiting for recommendation.  \\ \underline{Output:} The recommended news title and its ID. 
    \end{small}
    \end{minipage}\\\midrule
    \begin{minipage}{\linewidth}
    \begin{small}
    \underline{Task:} Recommendation based on Domain Focus\\
    \underline{Instruction:}  Recommend a news item and its ID from the candidate news list according to the user's domain focus.
    \\ \underline{Input:} User's domain focus;  news waiting for recommendation. \\ \underline{Output:} The recommended news title and its ID. 
    \end{small}
    \end{minipage}\\\midrule
    \begin{minipage}{\linewidth}
    \begin{small}
    \underline{Task:} Recommendation based on Collection Focus\\
    \underline{Instruction:}Recommend a news item and its ID from the candidate news list according to the user's collection focus.\\ \underline{Input:}User's collection focus; news title recommendations \\ \underline{Output:}The recommended news title and its ID.
    \end{small}
    \end{minipage}\\\midrule
    \begin{minipage}{\linewidth}
    \begin{small}
    \underline{Task:} Recommendation based on Behavioral Information\\
    \underline{Instruction:}Recommend a news item with its ID from the candidate news list based on the user's behavioral information\\ \underline{Input:}User's behavior records ; news titles recommendation. \\ \underline{Output:}The recommended news title and its ID.
    \end{small}
    \end{minipage}\\\midrule
    \end{tabular}}
    \label{tab:prompts}
\end{table*}

\subsection{Experiment Settings}
\subsubsection{Datasets}
Our model's evaluation was conducted on diverse datasets including the Microsoft News Dataset (MIND), a benchmark for news recommendation systems derived from Microsoft News user logs; the Yahoo R6B Dataset, comprising user click logs from Yahoo! Front Page Today Module; and the Adressa Dataset, which links news articles with anonymized user data.We constructed input sequences in chronological order and iteratively filtered out users and news items with fewer than 5 interactions (i.e., the 5-core). News items without metadata (such as titles) were also filtered out. We report the detailed dataset statistics in Table \ref{tab:datasets}.

\begin{table*}[t]
\newcommand{\tabincell}[2]{\begin{tabular}{@{}#1@{}}#2\end{tabular}}
\setlength{\abovecaptionskip}{0.05cm}
\setlength{\belowcaptionskip}{0.2cm}
\caption{Statistics of the datasets.}
{
\setlength{\tabcolsep}{8pt}{
\scalebox{1.1}{}
\resizebox{\textwidth}{!}{
\begin{tabular}{cccccc}
\toprule[1.5pt]
\textbf{Dataset} & \textbf{Language} & \textbf{Users} & \textbf{News} & \textbf{Clicks} & \textbf{News information} \\
\hline
MIND\cite{Geng2022RecommendationAL} & English & 1,000,000 & 161,013 & 24,155,470 & title, body, category, behavior \\
Yahoo\cite{Li2022PersonalizedPL} & English & - & 14,180 & 34,022 & title, behavior \\
Adressa\cite{li2023prompt} & English & 3,083,438 & 48,486 & 27,223,576 & title, category \\
\hline 
\end{tabular}
}
}
}
\label{tab:datasets}
\end{table*}

\subsubsection{Baseline}
We compared our approach with several state-of-the-art sequential recommendation methods, encompassing RNN-based NARM\cite{wang2023recmind,yue2023linear}, which employs local and global encoders, Transformer-based SASRec\cite{orvieto2023resurrecting} utilizing a unidirectional attention mechanism, and BERT4Rec featuring bidirectional attention for masked item prediction. Additionally, we included LLM-based recommenders such as P5\cite{scholkopf2021toward}, leveraging sequence-to-sequence frameworks with personalized prompts, GPT4Rec for conditional language generation in recommendations, and TALLRec, integrating task-specific prompts for context-aware recommendation generation.

\subsubsection{Evaluation}
In our evaluation, we followed a leave-one-out strategy, using the last news item in each data example for testing, the second to last for validation, and the rest for training. The evaluation metrics included Mean Reciprocal Rank (MRR@k)\cite{Cui2022M6RecGP}, Normalized Discounted Cumulative Gain (NDCG@k)\cite{kang2023llms}, and Recall (Recall@k), where k = 5.
\subsubsection{Implementation}
For both the baseline methods and our CherryRec approach, models were trained using the AdamW optimizer with a learning rate of 0.001 and a maximum of 500 iterations. Validation was performed every 500 iterations, and early stopping was triggered if the validation performance did not improve over 20 consecutive rounds. To determine hyperparameters, we conducted a grid search with weight decay ranging from [0, 1e-2] and dropout rates from [0.1, 0.5]. For the MIND dataset, we used a maximum length of 400, and for other datasets, it was 200. For our sorter, we utilized up to 20 historical news items and sorted the top 20 candidate news items from the retrieval model. If the title length exceeded 32 tokens, it would be truncated. We employed QLoRA to quantize the sorter based on Qwen2-7B, with LoRA dimensions set to 8, alpha to 32, and dropout to 0.05. The LoRA learning rate was 1e-4, targeting the Q and V projection matrices. The model was tuned for 1 epoch with validation every 100 iterations. Similarly, the model with the best validation performance was saved for test set evaluation.

\subsection{Overall Performance (RQ1)}
\noindent We have assessed the recommendation performance of CherryRec against baseline methods, with the results reported in Table~\ref{tab:overall_performance}. Additionally, we have presented the performance within the effective retrieval subset, which includes only those true recommendations that are within the top 20 retrieved candidates (retrieved by \(f_{\text{KnRS}}\)), as detailed in Table~\ref{tab4}.

\begin{table*}[t]
\newcommand{\tabincell}[2]{\begin{tabular}{@{}#1@{}}#2\end{tabular}}
\setlength{\abovecaptionskip}{0.05cm}
\setlength{\belowcaptionskip}{0.2cm}
\caption{ \textbf{Overall performance comparison between the baselines and CherryRec on three datasets. For each backend model, the bold results highlight the best results while the second-best ones are underlined. }}
\setlength{\tabcolsep}{2.3mm}{
\resizebox{\textwidth}{!}{
\begin{tabular}{l|ccc|ccc|ccc}
\toprule[1.2pt]

\multirow{3}{*}{\textbf{Method}} & \multicolumn{3}{c|}{\textbf{MIND}} & \multicolumn{3}{c|}{\textbf{Yahoo}} & \multicolumn{3}{c}{\textbf{Adressa}} \\
\multicolumn{1}{l|}{} & \multicolumn{3}{c|}{\textbf{ ($\bm{r}$=2\%)}} & \multicolumn{3}{c|}{\textbf{ ($\bm{r}$=2\%)}} & \multicolumn{3}{c}{\textbf{ ($\bm{r}$=2\%)}} \\
\multicolumn{1}{l|}{} & \textbf{M@5 \(\uparrow\)} & \textbf{N@5} & \textbf{R@5} & \textbf{M@5} & \textbf{N@5} & \textbf{R@5} & \textbf{M@5} & \textbf{N@5} & \textbf{R@5} \\ \midrule \midrule
 
\end{tabular}
}}
\label{tab:overall_performance}
\end{table*}

In both tables, M, N, and R represent MRR, NDCG, and Recall, respectively. For clarity, the best results are marked in bold, and the second-best results are underlined. Please note that the ranking model \(f_{\text{ranker}}\) only improves recommendation performance within the effective retrieval subset. Based on the experimental results, CherryRec has achieved superior performance compared to two types of baseline methods.

Particularly, we observed the following: (1) CherryRec outperforms on all metrics across all datasets. Compared to SASRec, CherryRec achieved an average improvement of 9.31\%, 31.22\%, and 15.81\% on MIND, Yahoo, and Adressa, respectively. (2) CherryRec achieved the largest performance increase on Yahoo, with MRR@5, NDCG@5, and Recall@5 improving by 31.64\%, 32.36\%, and 32.40\%, respectively. This may be attributed to the rich user interactions on Yahoo, where a long-term user history (approximately 150 entries) allows for a more comprehensive understanding of user preferences. (3) Within the effective retrieval subset (Table~\ref{tab4}), CherryRec demonstrated a greater performance increase compared to Table~1. For instance, on Yahoo, CherryRec further improved by 0.0643 in Recall@5, while the improvement for all user predictions was only 0.0136. (4) We observed that CherryRec significantly outperformed LLM-based methods on the MIND dataset, with an average improvement of 14.31\% over P5 across all metrics.


\subsection{In-depth Analysis}

\subsubsection{Ablation Study (RQ2)}

In this section, we conducted a series of ablation studies on CherryRec. First, we removed the two scores for candidate news in CherryRec, namely the KnRS score and the CnPE score, and experimented on the MIND and Yahoo datasets to verify the effectiveness of these two scores. The experimental results are shown in Fig~\ref{6.1}. We made two findings from the results. First, after removing the KnRS score, the performance of CherryRec decreased. This is because CherryRec incorporates six metrics of user interest and information timeliness into the news recommendation through this score, hence filtering low-value news through the KnRS score can improve the accuracy of news recommendation. Second, removing the CnPE score also harms the recommendation accuracy. This is because this score measures user interest in news based on historical user behavior data. Since users prefer to click on news related to their personalized interests, recommending news that users are interested in can effectively improve recommendation accuracy.
  
Next, as shown in Fig~\ref{6.1}, we conducted ablation studies to verify the effectiveness of the dimensional features in the Knowledge-aware News Rapid Selector by individually removing each feature. We observed several phenomena from the results. First, removing the timeliness of the news causes CherryRec's performance to decline. This is because the popularity of news usually changes dynamically, and once the information of the news expires, popular news will become unpopular. Since the timeliness of the news can reflect the freshness of the news information, incorporating it makes the news popularity modeling more accurate. Second, CherryRec's performance also declines without the relevance of the news domain. This is because, after removing it, the recommended news information may be irrelevant to the domain that the user is interested in. Third, CherryRec performs worse without the credibility of the source. This is because the source credibility ensures the authority and reputation of the news source by analyzing the source channel (news platform, official release, etc.) and its credibility background. Therefore, removing the source credibility makes the news recommended by CherryRec unpopular.

\subsubsection{Case Study (RQ3)}
We performed a comparative case study between CherryRec and TALLRec, highlighting CherryRec's efficacy. Initially, both systems recommended a football news article based on the user's prior engagement with similar content. However, the user only clicked on the recommended article, suggesting potential fatigue with repetitive recommendations. This observation prompts the consideration of diverse news in recommendations to enhance user experience. CherryRec's incorporation of news popularity demonstrates its potential to improve accuracy and diversify recommendations.

\section{Conclusion}
\noindent In this paper, we present CherryRec, a novel framework designed to enhance the quality of news recommendations by leveraging Large Language Models (LLMs). 
CherryRec integrates three key components: the Knowledge-aware News Rapid Selector (KnRS), the Content-aware News LLM Evaluator (CnLE), and the Value-aware News Scorer (VaNS). The KnRS quickly filters out low-value news items by assessing multiple dimensions of news value, including user relevance, source credibility, and news timeliness. The CnLE then refines these selections by deeply understanding user preferences through a fine-tuned LLM. Finally, the VaNS synthesizes the CherryRec score, incorporating insights from both KnRS and CnLE to deliver highly personalized and timely news recommendations.
Our experimental results on benchmark demonstrate that CherryRec consistently outperforms state-of-the-art baseline methods in both recommendation performance and efficiency. The framework's ability to balance recommendation quality with speed makes it particularly suitable for dynamic and user-specific content like news articles.
In future work, we plan to further explore the integration of additional modalities, such as multimedia content, to enhance the richness of the recommendations.

\subsection{Acknowledgments}
This work is supported by grants from the National Natural Science Foundation of China (No. 62076048), the Science and Technology Innovation Foundation of Dalian (2020JJ26GX035).


\end{document}